\documentclass[%
 aip,
 amsmath,amssymb,
 reprint,%
]{revtex4-1}

\usepackage{graphicx}
\usepackage{dcolumn}
\usepackage{bm}
\usepackage{hyperref}
\hypersetup{
    colorlinks=true,
    linkcolor=blue,
    filecolor=blue,
    citecolor=blue,
    urlcolor=blue,
}
\usepackage[utf8]{inputenc}
\usepackage[T1]{fontenc}
\usepackage{mathptmx}
\DeclareMathOperator{\sech}{sech}

\begin{document}

\preprint{AIP/123-QED}

\title{Terawatt attosecond X-ray source driven by a plasma accelerator}

\author{C. ~Emma}
\email{cemma@slac.stanford.edu.}
\affiliation{SLAC National Accelerator Laboratory, Menlo Park, California 94025, USA}
\author{X. Xu}
\affiliation{SLAC National Accelerator Laboratory, Menlo Park, California 94025, USA}
\author{A. Fisher}
\affiliation{University of California, Los Angeles, 90095, USA}
\author{J. P. MacArthur}    
\affiliation{SLAC National Accelerator Laboratory, Menlo Park, California 94025, USA}
\author{J. Cryan}
\affiliation{SLAC National Accelerator Laboratory, Menlo Park, California 94025, USA}
\author{M. J.~Hogan} 
\affiliation{SLAC National Accelerator Laboratory, Menlo Park, California 94025, USA}
\author{P. Musumeci}
\affiliation{University of California, Los Angeles, 90095, USA}
\author{G. White} 
\affiliation{SLAC National Accelerator Laboratory, Menlo Park, California 94025, USA}
\author{A. Marinelli}
\affiliation{SLAC National Accelerator Laboratory, Menlo Park, California 94025, USA}

\date{\today}

\begin{abstract}
Plasma accelerators can generate ultra high brightness electron beams which open the door to light sources with smaller physical footprint and properties un-achievable with conventional accelerator technology. In this paper we show that electron beams from Plasma WakeField Accelerators (PWFAs) can generate few-cycle coherent tunable soft X-ray pulses with TW peak power and a duration of tens of attoseconds, an order of magnitude more powerful, shorter and with better stability than state-of-the-art X-ray Free Electron Lasers (XFELs). Such a light source would significantly enhance the ability to experimentally investigate electron dynamics on ultrafast timescales, having broad-ranging impact across multiple scientific fields. Rather than starting from noise as in typical XFELs, the X-rays emission in this approach is driven by coherent radiation from a pre-bunched, near Mega Ampere (MA) current electron beam of attosecond duration. This relaxes the restrictive tolerances which have hindered progress towards utilizing plasma accelerators as coherent X-ray drivers thus far, presenting a new paradigm for advanced-accelerator light source applications.
\end{abstract}

\maketitle

\section{\label{sec:Introduction} Introduction}

Attosecond science is revolutionizing the way we understand and control electron motion at the quantum level \cite{Krausz2009,Corkum2007}. This revolution has been enabled by two types of attosecond light sources: ones based on High-Harmonic Generation (HHG) in gas \cite{McPherson,Ferray1988,Teichmann2016}, and more recently X-ray Free Electron Lasers (XFELs) \cite{Duris_XLEAP,Zhang2020}. HHG sources are capable of generating ultra-short pulses down to 43 as duration \cite{Gaumnitz2017}, with pulse energy in the pJ-range at soft X-rays. XFELs have the benefits of much larger pulse energy ($\sim$ 6 orders of magnitude) and wavelength tunability compared with HHG. These features enable experiments which investigate nonlinear electron dynamics \cite{ONeill2020} using X-ray spectroscopy methods including attosecond pump-probe techniques.  These experimental methods are very sensitive to the X-ray pulse intensity, and the generation of X-ray pulses with TW peak power would increase the amount of nonlinear X-ray interactions by one order of magnitude over state of the art XFELs. Furthermore XFEL pulses are currently limited to hundreds of attosecond pulse durations, with the shortest pulse recorded being 280 as \cite{Duris_XLEAP}. The temporal evolution of valence electronic excitations can be as fast 100 as, therefore even shorter X-ray pulses are highly desirable for the study of ultrafast electronic phenomena \cite{Krausz2009}. Finally, reaching below 100 as has interesting ramifications for probing the universal response of small systems to sudden removal of an electron (occurring on the 50-100 as timescale) and is critical for understanding correlated motion in atomic systems \cite{Breidbach2005}. 

The shortest possible pulse duration achievable via the FEL interaction $\Delta t_{min}$ is determined by the FEL gain bandwidth which scales with the gain length $L_G$ and the emittance $\epsilon$ of the lasing electron beam\cite{Saldin2004} $\Delta t_{min}\propto L_G\propto \epsilon^{5/6}$ . The lower bound on the available pulse duration is due to the finite slippage of the radiation with respect to the electron beam during an FEL gain-length. This problem can be mitigated by shortening the gain-length using high-current spikes generated with an infrared modulator (a method referred to as enhanced Self-Amplified-Spontaneous-Emission (SASE) \cite{ZholentseSASE}), or by enhancing the FEL bandwidth with a large energy-chirp \cite{Saldin2006}. The use of broadband beam instabilities \cite{SchneidmillerLSCInstability2010}  to generate short microbunched beams has also been proposed for the generation of VUV attosecond pulses \cite{dohlus2011generation, marinelli2013using}. Furthermore, Ref.  \cite{tibai2014proposal} proposed the use of an electron beam shorter than the resonant wavelength as a way to generate carrier-envelope-phase stable few-cycle pulses of UV radiation. Numerical studies of this method applied to state of the art photo-injector beams as well as laser-plasma accelerators suggest that the pulse energy available from this method is in the nJ range \cite{tibai2018carrier}. Finally, in the context of advanced accelerator light sources, Ref. \cite{Alotaibi2020} numerically examined the effects of coherent emission from ballistically bunched electron beams using a plasma photocathode, obtaining sub-$\mu$J pulse energies from fs-long pulses at UV wavelengths.

\begin{figure*}
\includegraphics[width=\linewidth]{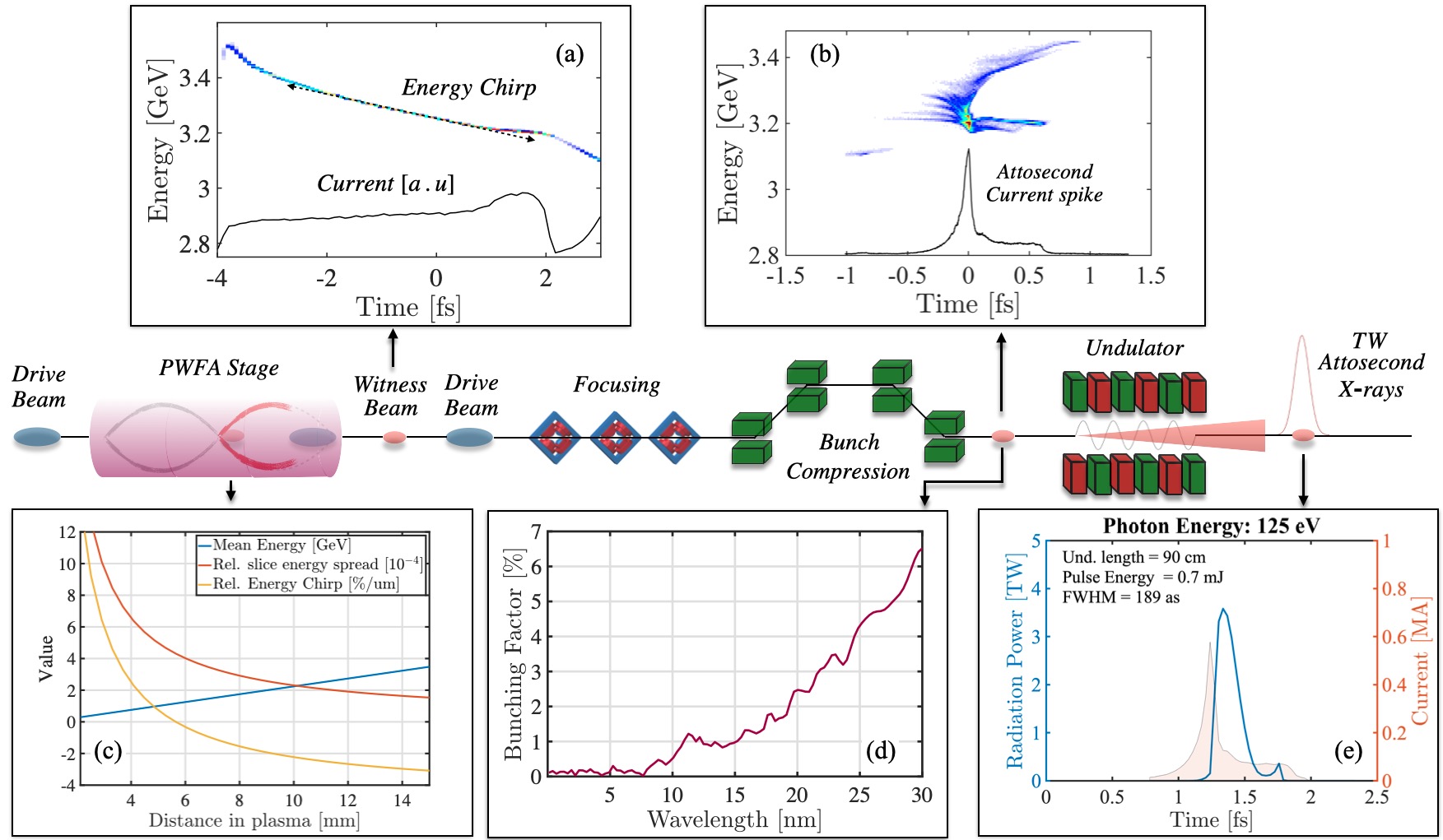}
\caption{Schematic of a PWFA-driven attosecond X-ray source based on coherent radiation from an attosecond long electron beam. (a)-(c) The PWFA stage imparts a large energy chirp in the few fs core of the bunch which is used to fully compress the beam in a weak chicane achieving near MA current with $\%$-level energy spread. (d)-(e) The beam's $\%$-level bunching factor at X-ray wavelengths enables high power coherent emission in a m-length undulator. This approach relaxes the requirement on beam emittance, energy spread and pointing stability which have thus far hindered the realization of a high-gain FEL driven by a plasma accelerator.} 
\label{Fig_1}
\end{figure*}

In this article we show that few-cycle coherent soft X-ray pulses in the 0.1-1 mJ energy range with sub-100 as pulse duration can be obtained using beams generated by a Plasma Wakefield Accelerator (PWFA), thus combining the benefits of HHG and XFEL sources. Rather than employing the high-gain SASE FEL principle starting from noise \cite{Bonifacio1984}, the X-ray emission in this source is driven by coherent radiation from a pre-bunched, near Mega-Ampere (MA) current electron beam of attosecond duration (see Fig.\ref{Fig_1}) \cite{Doria1993,Jaroszynski1993,Krinsky1999,Piovella1999,McNeil1999,Huang2000,Gover2019}. This approach presents a new paradigm for PWFA driven light sources as it relaxes the tight beam energy spread, emittance and pointing stability constraints which have hindered progress towards achieving XFEL gain thus far. Finally, this enables the generation of tunable TW-attosecond pulse pairs with attosecond timing jitter, with application to X-ray pump-probe experiments. We note that this approach is applicable to laser-driven wakefield accelerators where beams with similar properties can be generated at the exit of the plasma stage, albeit with typically lower charge \cite{EsareyRMP}. We present simulations of the scheme, comparing the photon beam properties with state-of the art attosecond X-ray sources. 

Using plasma-based accelerators as coherent X-ray light source drivers has long been a goal of the accelerator community \cite{Grner2007,Nakajima2008,Couprie2014,vanTilborg2017}. To date, the challenging longitudinal and transverse beam parameters required to drive an XFEL have made this a prohibitive task. To generate XFEL gain the beam's relative energy spread must be smaller than the Pierce parameter\cite{Bonifacio1984} $\rho$  (typically $\mathcal{O}(10^{-3})$ for state of the art XFELs), while in the transverse dimension the geometric emittance must satisfy the Pellegrini criterion $\epsilon \lesssim \lambda/4\pi$, where $\lambda$ is the radiation wavelength \cite{Kim1985,Pellegrini1990}. Furthermore, the beam trajectory inside the undulator must be accurately controlled to avoid degradation of the X-ray power. The angular misalignment tolerance is characterized by the critical angle $\theta_c = \sqrt{\lambda/L_G}$, and sets the tolerance to $<$ 10 $\mu$rad for XFELs \cite{Tanaka2004, Huang2007}. A recent estimate has concluded this tolerance is $>$18x tighter than achievable for existing PWFA facilities \cite{White2019}.

Adding to these stringent requirements, one further challenge facing FEL applications of plasma accelerators is the presence of large energy time-energy correlations (chirps) in the accelerated electron beams, which hinders the FEL gain process. This can be mitigated by shaping the current of the trailing witness bunch to flatten the wakefield \cite{Bane1985, Tzoufras2008} but remains challenging for injection scenarios in which the witness beam is injected and accelerated inside the plasma. Recent studies have also explored using multiple plasma stages \cite{FerranPousaPRL,dArcy2019} or accelerating multiple bunches in a single stage \cite{Manahan2017} to flatten the electron beam chirp at the undulator entrance. Finally, other proposed solutions rely on stretching the chirped electron beam longitudinally or transversely to achieve a sufficiently small slice energy spread \cite{Maier2012,Huang012}. These methods offer a path forward towards a fifth-generation light source, however, with the exception of Ref. \cite{Manahan2017}, they intrinsically trade off by reducing the beam brightness to enable the lasing process. This foregoes one of the main advantages of plasma based accelerators when compared to conventional RF systems, which enables performance gains for the light source beyond reducing the physical footprint of the accelerator facility. 

\begin{figure}[t]
\includegraphics[width=\linewidth]{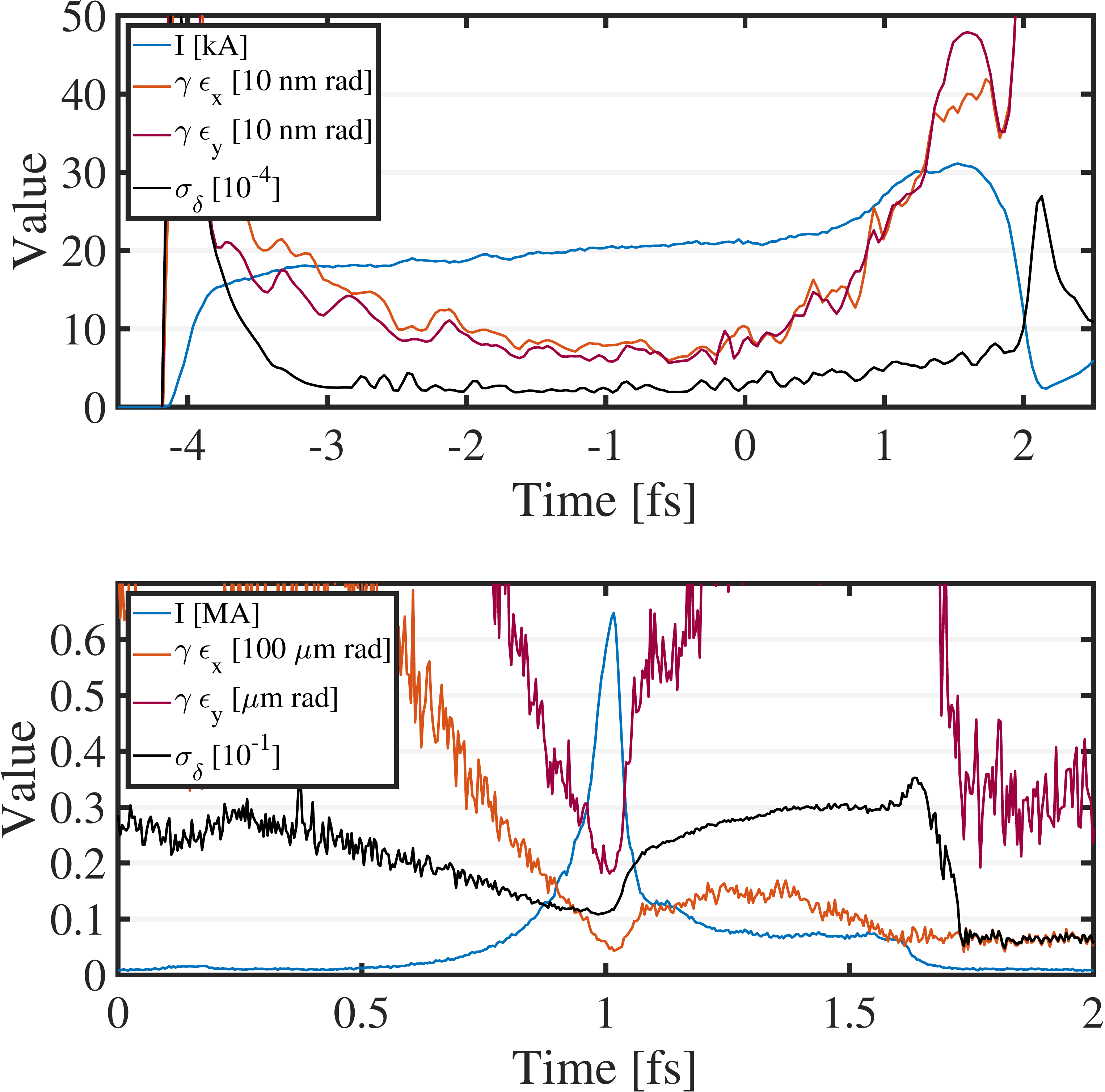}
\caption{Electron beam slice parameters at the plasma exit (top) and at the undulator entrance (bottom) corresponding to the longitudinal phase spaces shown in Fig. 1 a,b.} 
\label{Fig_2}
\end{figure}

\section{Electron beam dynamics of the PWFA-driven attosecond X-ray source}

The approach we present in this work leverages the ultra-high longitudinal brightness and strong chirps naturally present in PWFA beams to further compress the bunch after the exit of the plasma to a single current spike of attosecond duration (see Fig. \ref{Fig_1}). In a similar vein to enhanced SASE, we rely on a weak bunch compressor to avoid excessively degrading the beam brightness through Coherent Synchrotron Radiation (CSR). Unlike enhanced SASE, however, here we compress the electron beam to  a duration comparable to the final radiation wavelength. We demonstrate our approach using the same plasma and electron drive beam parameters as Ref. \cite{Xu2017}, where it was shown that electron beams can be injected and accelerated to GeV energy with $\sim 10^{-4}$ slice energy spread and bunch charge in the 100 pC range using the density downramp mechanism \cite{Bulanov1998,Suk2001,Ullmann2020}. We adopt the same density downramp with an elongated accelerating section compared to Ref. \cite{Xu2017} (which stops at an energy of 1.61 GeV), scaling the chirp and final energy to 3.3 GeV (see Table \ref{Table1} and Fig. \ref{Fig_1}), having chosen this final energy to facilitate comparison between our scheme and state of the art attosecond XFELs \cite{Duris_XLEAP,Zhang2020}. 
\begin{table}
\caption{\label{Table1}Simulation parameters for the PWFA-driven TW-attosecond X-ray source.}
\begin{ruledtabular}
\begin{tabular}{lcr}
\textbf{Parameter} & \textbf{Value} & \textbf{Unit}\\
\hline\hline
\textbf{Plasma and Drive Beam} & \\
Plasma Density &  1.1$\rightarrow$ 1.0$\times$ 10$^{18}$  & cm$^{-3}$\\
Downramp Length & 100 & $c/\omega_p$\\
Beam Energy &   2 &  GeV \\
RMS beam size (r,z) &   (2.7,5.3) & $\mu$m \\
Peak Current &   34 & kA \\
\hline
\textbf{Triplet and Chicane} & \\
Quadrupole strengths &  -48.8,15.3,-7.8  &  m$^{-2}$ \\
Dipole Bend Angle &   3.4 &  mrad \\
Momentum Compaction &  27 &  $\mu$m\\
\hline
\textbf{Witness Beam} (at Undulator) & \\
Beam Energy &   3.3 &  GeV \\
Peak Current &  0.65 &  MA\\
Norm. RMS Emittance (x,y) &   (4.3 , 0.18) &  $\mu$m rad\\
FWHM Beam Size (x,y,z) &  (10.8, 7.5, 0.023) & $\mu$m \\
RMS Slice Energy Spread &  1.4 & $\%$ \\
\hline
\textbf{Undulator} & \\
RMS Undulator Parameter &   3.73 & -\\
Undulator Period &  5.6 &  cm \\
Number of Periods &  20 & -\\
\hline
\textbf{Radiation} & \\
Wavelength &    10 & nm\\
Peak Power &    0.25-3.8 & TW\\
FWHM Pulse Duration &   38-294 & as\\
\end{tabular}
\end{ruledtabular}
\end{table}

\begin{figure*}
\includegraphics[width=\linewidth]{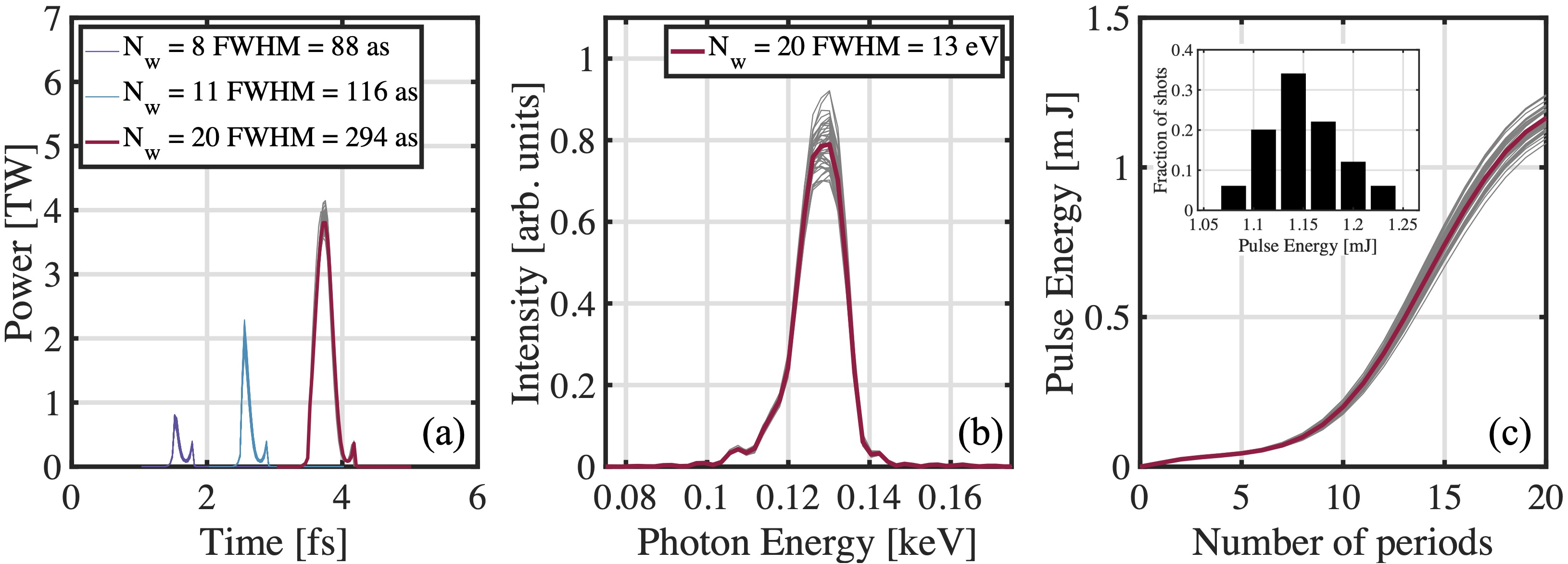}
\caption{X-ray simulation results for the PWFA-driven attosecond X-ray source. (a)-(c) Power, spectrum and pulse energy from 50 independent $\texttt{GENESIS}$ runs with distinct random seeds for the initialization of the particle co-ordinates and the field. The pulse length and spectral bandwidth exhibit high reproducibility with a mean ($\pm$ rms) pulse energy at the undulator exit of $1.2 \pm 0.038$ mJ.} 
\label{Fig_3}
\end{figure*}

The downramp injection process for this set of parameters has been explored in detail in Ref. \cite{Xu2017} and here we focus our discussion on the salient beam dynamics occurring in the accelerating section following the downramp and prior to the plasma exit. The simulations of the PWFA-stage are performed using the 3D PIC code $\texttt{OSIRIS}$ \cite{Fonseca2002}. Simulation results and fits of the witness beam energy, linear chirp and energy spread in the accelerating section are shown in Fig.\ref{Fig_1}c. In the nonlinear blowout regime \cite{Lu2006}, the witness beam is accelerated by an average field $E_z$ with an approximately constant linear field gradient along the bunch $E_z^{\prime}\approx m  \omega_p^2/2e$, where $\omega_p^2= e^2n_p/m\epsilon_0$ is the plasma frequency and $n_p$ is the plasma density. Electrons from the edge of the plasma sheath are injected and trapped towards the back of the bubble where
$E_z\approx (mc\omega_p/e) \left\langle\xi/2\right\rangle=$ 0.21 TV/m and $\left\langle\xi\right\rangle=4.3$ is the average position of the injected electron beam relative to the center of the bubble in units of the plasma skin depth $c/\omega_p$. For a sufficiently long accelerating section the witness beam chirp $h\equiv \gamma^{-1} d\gamma/dz$ approaches an asymptotic value $h_\infty = E_z^{\prime}/E_z\propto \omega_p \approx - 3.7\%/\mu$m for our parameters (see Fig. \ref{Fig_1}c). The variation of the chirp due e.g. to variations in the plasma density or the length of the accelerating section \cite{Zhang2019}, is reduced in the asymptotic limit. This reduces the fluctuations of the bunch length after compression and consequently improves the stability of the X-ray pulses whose properties are determined by the incoming electron beam profile. This is in contrast to typical SASE XFELs where amplification of chaotic noise in the electron beam results in well-known statistical fluctuations of the X-ray pulse properties \cite{BonifacioFluctuations}. 

The transport of the beam from the plasma to the undulator entrance is composed of a quadrupole triplet and a four-dipole chicane and is modeled with the particle tracking code $\texttt{Lucretia}$ \cite{Tenenbaum2005} including the effects of CSR and Longitudinal Space Charge (LSC) on the particle distribution. The resulting current profile, slice energy spread and emittance before and after transport and compression are shown in Fig \ref{Fig_2}. The triplet is used to focus the divergent beam from the exit of the PWFA to the undulator entrance. To first order, the chicane compresses the electron beam to a final bunch length $\sigma_{z,f}^2=(1+hR_{56})^2\sigma_{z,i}^2+R_{56}^2\sigma_{\gamma,i}^2$ where $R_{56}$ is the momentum compaction factor and $\sigma_{(z,\gamma),i}$ are the incoming bunch length and energy spread respectively. Maximum compression is achieved when $R_{56}= |h|^{-1}\approx |h_{\infty}|^{-1}=$ 27$\mu$m  for our parameters. The minimum bunch length is limited by the incoming slice energy spread to $\sigma_{z,f}=R_{56}\sigma_{\gamma,i}\approx 6$ nm. This underscores the importance of maintaining a small slice energy spread while simultaneously imparting a large energy chirp to enable coherent emission at short wavelengths. 

Nonlinear compression, LSC and CSR effects modify the electron beam phase space during the compression process from the simplified linear picture. The CSR emitted in the compressor dipoles is the dominant of these effects and causes the energy to change along the beam, increasing the angular spread and hence the emittance of the compressed bunch. In the thin beam approximation ($\sigma_r R_{b}^{-1/3}\sigma_z^{-2/3} < 1$, where $R_b$ is the dipole bending radius) the impact of these effects is adequately described by a 1D model  \cite{Derbenev1995,SALDINCSR,BorlandCSR}. The large energy chirp leads to a small $R_{56}$ and small bend angle which is advantageous for suppressing CSR effects. We adopt the prescription of small $\beta$ function at the entrance of the final dipole to minimize the CSR induced emittance growth \cite{Dohlus2005,DohlusICFA,DIMITRI20141}. The emittance growth is balanced by the increase in peak current through compression and as shown in the X-ray simulations in the following section, the coherent emission process is not critically hindered by this effect. The current profile exhibits an asymmetric 0.64 MA-current spike of length $c\Delta t$ = 23 nm FWHM. The magnitude of the bunching factor $b = |\left\langle e^{2\pi i ct_j/\lambda}\right\rangle|$ evaluated at $\lambda = 10$nm is 0.5 $\%$  (see Fig. \ref{Fig_1}d,\ref{Fig_2}) and drives the emission of radiation in an ultra-short (m-length) downstream undulator.

\section{PWFA-driven attosecond X-ray source properties}

The X-ray simulations (see Table \ref{Table1} for parameters) are performed using the 3D code $\texttt{GENESIS}$ 1.3 which uses the slowly varying envelope approximation and a period averaged current to compute and propagate the radiation field discretized on a Cartesian grid.  These approximations, along with a comparison with similar results obtained using GPTFEL \cite{fisher2020} a non period averaged frequency based code, are discussed further in the Appendix. The stability of the power profile, spectrum and pulse energy are shown in Fig. \ref{Fig_3} where we display results from 50 simulations with different random noise initialization for the particles and the fields. The FWHM pulse duration evolves from 36-238 as and the peak power grows from 0.25-3.8 TW while the beam traverses the 20 period undulator (see Fig.\ref{Fig_3}a). The FWHM time-bandwidth product 3.8 eV$\cdot$fs, close to a factor of 2 from a Fourier limited pulse. The pulse energy ($\pm$ rms) at the undulator exit is is $1.2 \pm 0.038$ mJ, one order of magnitude larger than state-of-the art attosecond SASE-XFELs, with an order of magnitude smaller fluctuations (see Fig.\ref{Fig_3}c inset) as a result of the radiation being driven by coherent emission from a pre-bunched electron beam. The high peak current in the compressed bunch causes a $\%-$level energy chirp to develop as a result of LSC in the undulator \cite{Ding2009}. This is included in the simulation and can be compensated by an appropriate taper of the undulator magnetic field \cite{Saldin2006}. The pulse length and peak power can be tuned by changing the number of undulator periods allowing experiments to choose the X-ray pulse properties (pulse length and peak power) as desired for different applications. 
Alternatively, the pulse length can be tuned on-the-fly by adjusting the magnetic field taper amplitude, and in our simulations a stronger linear taper generates shorter pulses, at the expense of smaller peak power.

\begin{figure}
\includegraphics[width=\linewidth]{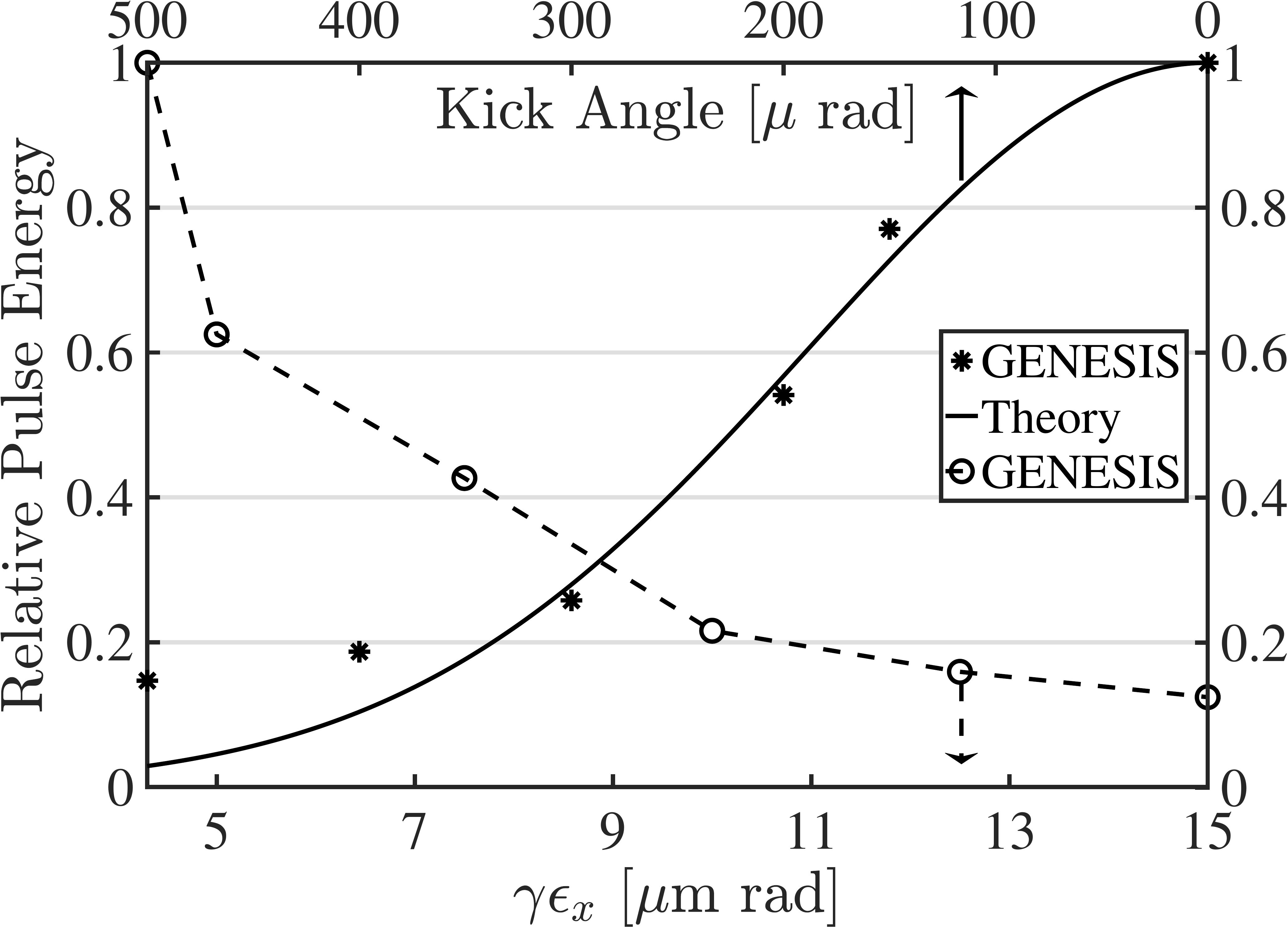}
\caption{Relative X-ray pulse energy for increasing angular kicks or increased emittance of the electron beam at the undulator entrance. The tolerance to beam emittance is up to 15 $\mu$m rad and the tolerance to angular kicks up to 500 $\mu$rad, more than 1 order of magnitude less restrictive than typical for XFELs.} 
\label{Fig_4}
\end{figure}

As a critical step in evaluating the feasibility of our approach we have quantified the tolerances of the coherent X-ray emission process to beam emittance and pointing jitter. The impact of beam emittance was evaluated by artificially increasing the beam horizontal (bend plane) emittance at the undulator entrance in the range $\gamma \epsilon_x =$ 4.3-15 $\mu$m rad (see Fig. \ref{Fig_4}). The relative X-ray pulse energy falls below 50 $\%$ for values of the emittance larger than $\sim 7 \mu$m rad, a value 40 $\%$ larger than the limit set by the Pellegrini criterion for our parameters. The tolerance to pointing errors was similarly investigated by artificially introducing angular beam misalignments at the entrance of the undulator. The $\%$-level bunching allows the beam to radiate strongly for large incoming kick error up to 500 $\mu$rad, an improvement of 1-2 orders of magnitude compared to with typical XFEL tolerances in the $\sim $1-10 $\mu$rad range \cite{Tanaka2004,Huang2007}. The simulations results are in reasonable agreement with theoretical estimates of power degradation from a kicked electron beam (see eq. 4 in Ref. \cite{Geloni2018}). Such a relaxed angular misalignment tolerance may enable multiplexing techniques \cite{MacArthur2018} for single or two-color attosecond pulse generation and find applications in attosecond pump-probe experiments. 
\section{Conclusion}
In conclusion we have shown that electron beams from PWFAs can be used to generate tunable few-cycle TW-power, sub-100 attosecond long X-ray pulses, an order of magnitude more powerful, shorter and with better stability than possible with state of the art XFELs. A novel light source based on this approach would rival HHG sources in the shortest possible pulse durations while increasing the power and maintaining the flexibility of XFELs, opening up new science in the study of electron motion in atomic and molecular systems. 

Of critical importance for PWFA-driven light source applications, the approach we presented is also an order of magnitude less sensitive to emittance, energy spread and angular trajectory jitter inside the undulator compared to high-gain SASE FELs starting from noise \cite{White2019}. This is due to a combination of the very high peak current and the ultra-short undulator length. As a result this substantially relaxes the requirements for preservation of electron beam quality during transport from plasma to undulator which have hindered progress in utilizing advanced accelerators as drivers for XFELs thus far. We note that scaling this method to shorter wavelength may be possible by accelerating the witness beam to higher energy ($\sim$5-10 GeV) using a higher-energy drive beam and a longer plasma stage. Finally, strong interest in the generation of ultra-short high intensity electron beams with MA peak current was also raised by recent proposals for high-field colliders \cite{White2018,HEPGARD2019} and for studying non-linear quantum electrodynamics processes \cite{Yakimenko2019}.  The development of such a light source would enable studies of beam dynamics at the frontier of high-brightness electron and photon science, a topic of high interest for the accelerator physics community and beyond. 
\begin{acknowledgments}
Work supported by the U.S. Department of Energy under contract number DE-AC02-76SF00515. This work was also partially supported by DOE grant DE-SC0009914. The OSIRIS simulations were performed on the National Energy Research Scientific Computing Center (NERSC).
\end{acknowledgments}

\appendix*

\section{X-ray simulation methods}

We discuss details of the approximations made in the X-ray simulation using $\texttt{GENESIS}$ 1.3 and compare these with calculations using 1D FEL theory, a 1D FEL code and a non-period averaged 3D code $\texttt{GPTFEL}$ \cite{fisher2020}. Two important approximations $\texttt{GENESIS}$ uses to solve the Maxwell equations and calculate the radiation field are the Slowly Varying Envelope Approximation (SVEA) and the period average current approximation. These two approximations result in two requirements on the electron beam and radiation field: 

\begin{enumerate}
    \item $|dE/dz|\ll k|E(z)|$
    \item $\lambda \le \Delta s \ll l_{c}$
\end{enumerate}

where $s= z-ct$ is the longitudinal bunch co-ordinate and $l_{c} = (\lambda/\lambda_w) L_G$ is the FEL cooperation length. The SVEA (1) requires that the field does not change appreciably on the spatial scale of one wavelength while the period average approximation (2) requires that the field and current are sampled more frequently than the characteristic time over which the field changes $l_c/c$. In our simulations we have $L_G\sim 3\lambda_w$, $ l_c\sim$ 3$\lambda$, and we employ the minimum sampling interval of $\Delta s = \lambda$ to resolve the electron beam and radiation field. To avoid instabilities of the field solver the field integration step-size $\Delta z$ must be smaller than the characteristic spatial-scale for large changes in the field (the local gain length) which in our simulations is $\Delta z/L_G(z)\geq 3$. Furthermore, $\texttt{GENESIS}$ 1.3 assumes the current profile remains constant along the undulator, discretizing the electron beam in slices and preventing particles from moving between them. This approximation breaks down if the energy spread of the beam reaches very large values $\sigma_\gamma \sim 2/N_w$ but is not the case in our simulations. Finally, we include LSC effects in the simulations using a wakefield model (e.g. see eq. 1 in Ref. \cite{Ding2009})  while $\texttt{GENESIS}$ does not calculate the effect of transverse space charge on the electron beam. 

In order to evaluate the impact of these approximations, we compare the results of $\texttt{GENESIS}$ with $\texttt{GPTFEL}$ \cite{fisher2020} a non-period averaged frequency based code built on the General Particle Tracer tracking code \cite{DeLoos1996} which self-consistently includes the effects of longitudinal and transverse space charge on the electron beam. The results of such a comparison are shown in Fig. 5. Both simulation codes confirm the unique features of this radiation source: TW peak power pulses with sub-100 attosecond pulse duration in a very short undulator. We see that in both cases the radiated pulse energy saturates after $\Delta z \sim 6 L_G$, consistent with an exponentially growing emission starting with an initial bunching factor of $|b| = 0.5\%$. Some of the differences between the two simulation results are the slower startup evidenced in the first 5 periods, the shorter gain length ($L_G = 2.6 \lambda_w$) and the shorter final pulse length in the $\texttt{GPTFEL}$ simulation. These differences can be attributed to the a finer resolution of the current profile and the electron beam micro-structure in $\texttt{GPTFEL}$ which gives a more accurate evaluation of the initial bunching factor seeding the radiative interaction and its evolution along the undulator. The shorter final pulse length can be explained by the absence of the period-averaged current approximation in $\texttt{GPTFEL}$ which allows the code to resolve finer structure in the radiation field (with a sampling interval $c\Delta t = \lambda/30$) and higher frequency content outside the bandwidth of the $\texttt{GENESIS}$ simulation. We note that recent work \cite{Campbell2020} has performed a comparative analysis of ultra-short bunch simulations from a SVEA code similar to $\texttt{GENESIS}$ with a particle-in-cell code which integrates the Maxwell's equations without approximation \cite{Freund2017,Campbell2012}. The results report agreement between the two codes and experimental data for bunch lengths comparable to the co-operation length, as is the case for our parameters. Previous studies have also shown that when SVEA codes disagree with PIC or un-averaged codes for ultra-short bunches they tend to give conservative estimates of the radiation output, and the peak power and pulse length may be further enhanced and shortened by coherent effects not captured in SVEA codes \cite{Piovella1999,McNeil1999}. 

\begin{figure}[t]
\includegraphics[width=\linewidth]{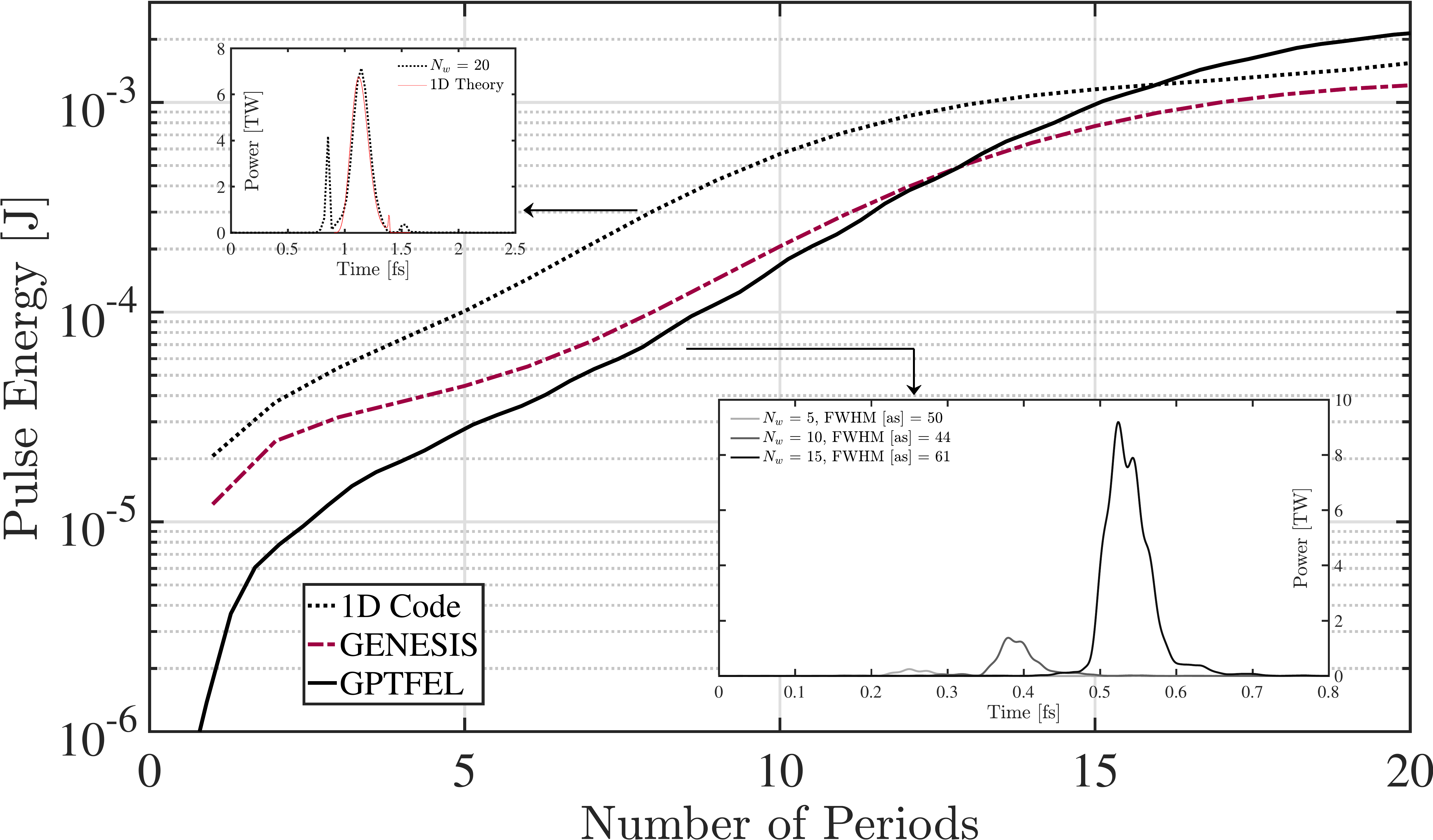}
\caption{Comparison of simulation results for the plasma-driven attosecond X-ray source using a 1D FEL code, $\texttt{GENESIS}$ and $\texttt{GPTFEL}$. The comparison confirms the source properties with all three codes: TW power pulses with sub-100 as pulse duration in an ultra-short undulator.} 
\label{Fig_1Supp}
\end{figure}

Finally, we compare the results of our simulations with the well-established 1D model of the `superradiant' regime of the FEL, which describes radiative emission in situations where the bunch length is comparable to the cooperation length (see e.g. Ref. \cite{Bonifacio1988,Bonifacio1990,Piovella1991}). A closed form solution for the normalized radiation field intensity $|A|^2$ in the leading radiation pulse can be derived in this limit and is based on the hyperbolic secant function:

\begin{equation}
    |A|^2 = \frac{z_1^2}{y}\sech^2\left[\frac{3\sqrt{3}}{2}\left(\frac{y}{2}\right)^{2/3}+ \log \left(\frac{|b_0|}{\sqrt{12\pi y}}\right)\right]
\end{equation}

where $|b_0|$ is the initial bunching factor, $z_1$ is the longitudinal co-ordinate along the bunch in units of the co-operation length, $y = \sqrt z_1(z/L_G-z_1)$ and the power in MKS units is given by $P = 4\pi A_b\rho\gamma m_e c^2/Z_0$ where $A_b$ is the electron beam cross-sectional area. We plot this result using our electron beam parameters and compare this with numerical integration using the same electron beam in a period-averaged 1D FEL code (see Fig. 1 top-left inset) \cite{Emma2017}. The 1D code models the effect of electron beam emittance as an additional energy spread and incorporates the energy modulation due to LSC by calculating the LSC wake function and applying the corresponding energy change to the electrons every undulator period. We find that eq. 1 agrees well with the results of 1D numerical integration for radiation in the leading pulse. Similarly to Ref. \cite{Bonifacio1990}, we observe an adjacent radiation spike in the 1D numerical results which is the result of bunching structure that persists in the electron beam when the leading radiation pulse has slipped out of the current spike. In the 1D analysis we see the saturation length is shortened compared to 3D simulation results as the 1D microscopic FEL equations neglect transverse effects which reduce the coupling between between electron beam and radiation pulse and drive the coherent emission process. For example, the 1D theory assumes the transverse electron beam spot-size is much larger than the radiation mode size $\sigma_x \gg \sigma_r = \sqrt{\lambda L_G/4\pi}$, a condition which is violated in our simulations where we have $\sigma_x \sim \sigma_r$. We note that the properties of the leading radiation pulse obtained in using this simplified 1D analysis are in reasonable agreement with those obtained from fully 3D simulations. Confirmation of the established 1D power scaling provides a simple and convenient way of estimating the radiation pulse properties at different wavelengths as a function of the electron beam parameters and the initial bunching factor.

\nocite{*}
\bibliography{main}

\end{document}